\documentclass[aps,prl,preprint,groupedaddress,showpacs]{revtex4}
\usepackage[dvips]{graphicx}
\begin{document}
\title{Fluctuations Do Matter: Large Noise-Enhanced Halos in Charged-Particle Beams}
\author{Courtlandt L. Bohn$^{1,2}$ and Ioannis V. Sideris$^1$}
\affiliation{$^1$Northern Illinois University, DeKalb, IL 60115
\\
$^2$Fermilab, Batavia, IL 60115}
\date{\today}
\begin{abstract}
The formation of beam halos has customarily been described in terms of a particle-core model in which the space-charge field of the oscillating core drives particles to large amplitudes.
This model involves parametric resonance and predicts a hard upper bound to the orbital amplitude of the halo particles.
We show that the presence of colored noise due to space-charge fluctuations and/or machine imperfections can eject particles to much larger amplitudes than would be inferred from parametric resonance alone.
\end{abstract}
\maketitle

Beam loss is a major concern for high-current light-ion accelerators such as are needed to drive high-intensity spallation neutron sources.
Just a tiny impingement, $\sim$1 nA m$^{-1}$GeV$^{-1}$, could generate radioactivation that would preclude routine, hands-on maintenance~\cite{jameson96}.
For a 1 mA, 1 GeV light-ion beam, i.e., for baseline beam parameters of the Spallation Neutron Source (SNS) presently under construction~\cite{sns03}, this criterion translates to just 1 in $10^6$ particles lost per meter, a quantity that scales linearly with average beam current.
Accordingly, a comprehensive understanding of beam-halo formation is imperative.

Early efforts to identify the fundamental mechanisms of halo formation centered on the use of a `particle-core' model~\cite{yjchen,oconnell93,gluckstern94}.
The basic recognition was that if a uniform-density core is made to pulsate, particles that initially lay outside the core and that resonate with its pulsations could reach large amplitudes and form a `halo'.
This led to the identification of parametric resonance as the essential mechanism of halo formation.
A key feature of parametric resonance in the context of the particle-core model is a hard upper bound to the amplitude that a halo particle can reach~\cite{oconnell93}.
Because the particle's orbital frequency is a function of its amplitude, at sufficiently large amplitude the particle falls out of resonance with the core and thereby its amplitude ceases from growing further.
The prospect that the beam halo is `self-collimating' has led to hope that aperture requirements for beamline components might be modest.
Smaller apertures are preferred in that, for example, they favor higher-efficiency operation of the accelerating cavities.
In turn, a wealth of studies and a large body of literature has developed over the past ten years that has centered on deciphering the maximum halo amplitude.
Refs.~\cite{okamoto97,gluckstern98,ikegami99,ikegami299,qiang00,wang00,jeon02,allen02} constitute a small sample; Ref.~\cite{allen02} documents a recent halo experiment and alludes to a maximum amplitude, though that amplitude could not be measured.

An urgent question is whether there is any physics not included in the particle-core model that could significantly influence the maximum particle amplitude.
One feature that is unavoidable in real accelerators but is commonly overlooked in simulations is the presence of noise.
The noise will manifest itself by way of the electromagnetic fields external to the beam, which then self-consistently influence the beam's evolving space-charge potential.
Noise sources could include hardware irregularities that establish fluctuating image-charge forces, jitter in power supplies, misalignments and/or asymmetries of beamline components, etc.
In the context of simulations, it could also include details in the space-charge potential that the simulation cannot model precisely.
A charged particle will experience all of the noise inherent to the total potential.
Moreover, the noise will generally comprise a superposition of `colored' noise, i.e., that for which the autocorrelation time is nonzero.
For example, the autocorrelation time of noise in the collective space-charge potential could be short, say of the order of a plasma period, whereas for hardware irregularities/misalignments it could be long, say several betatron (orbital) periods.
Herein, by generalizing simple particle-core models to include noise, we show that the presence of colored noise can potentially boost statistically rare particles to ever-growing amplitudes by continually kicking them back into phase with the core oscillation.

Following the ground-breaking work that introduced the particle-core model~\cite{oconnell93,gluckstern94}, we consider particles on radial orbits through an infinitely long, axially symmetric, uniform-density beam `core' that pulsates at a single frequency due to an imbalance, i.e., mismatch, between the repulsive, collective space-charge force and the confining external focusing force.
The core radius oscillates according to an equation of motion for the beam envelope.
Upon linearizing the envelope equation in terms of the core-oscillation amplitude, one finds the solution $R(t)=R\left[1+(M-1)\cos\omega t\right]$, wherein $\omega$ is the core-oscillation angular frequency and $M=R(0)/R$ is the mismatch parameter, i.e., the ratio of initial-to-matched core radii.
For the uniform-density core, i.e., zero-temperature beam, the core-oscillation frequency is $\omega=\sqrt{2}\Omega$, where $\Omega$ denotes the external focusing angular frequency.
The particle orbits are governed by the dimensionless equation of motion
\begin{equation}
\ddot{x}+x\left[1-{{\Theta(1-|x|)}\over{\left[1+(M-1)\cos\omega t\right]^2}}-{{\Theta(|x|-1)}\over{x^2}}\right]=0;
\label{eq:wangler}
\end{equation}
the transverse coordinate $x$ is normalized to the radius $R$ of the matched, hence stationary, beam; time is multiplied by $\Omega$ which means all frequencies are expressed as multiples of $\Omega$; and $\Theta(u)$ is the Heaviside step function.
The second and third terms in square brackets govern the motion of the particle when it is inside and outside the core, respectively.
This model will henceforth be called ``Model I''.

Because Model I is strictly one-dimensional and contains a discontinuity in the form of a step function, we shall also study a second model for which the unperturbed beam is a spherically symmetric configuration of thermal equilibrium (TE) computed in Ref.~\cite{bohn03}.
The dimensionless equation of motion for this model, henceforth called ``Model II'', is
\begin{eqnarray}
\ddot{{\bf x}}&=&-{\bf\nabla}\Psi;\;\;
\Psi=\Psi_0+\Psi_1;
\nonumber\\
\Psi_0&=&\frac{1}{2}\Omega^2r^2+\Phi(r),\;\;\Psi_1=\mu\Phi(r_1)\sin\omega t,
\nonumber\\
r&=&\sqrt{x^2+y^2+z^2},\;\;r_1=\sqrt{0.8(x^2+y^2)+z^2};
\label{eq:TE}
\end{eqnarray}
in which the external focusing angular frequency is $\Omega=1.0001/\sqrt{3}$.
As explained in Ref.~\cite{bohn03}, the coordinates and time are measured in units of Debye length and inverse plasma angular frequency, so the normalization differs from that of Model I.
The self-potential $\Phi(r)$ corresponds to ``intermediate space charge''; the associated density drops with radius over a length scale comparable to that of the quasi-uniform core.
The potential $\Psi_1$ is a prolate spheroidal perturbation whose strength corresponds to the parameter $\mu$.

To Models I and II we add fluctuations in the form of Gaussian colored noise such that $\omega\rightarrow\omega(t)=\omega_0+\delta\omega(t)$, with $\delta\omega(t)$ sampling an Ornstein-Uhlenbeck process.
Its first two moments fully determine the statistical properties of the noise:
\begin{equation}
\langle\delta\omega(t)\rangle=0,\;\;\;\langle\delta\omega(t)\delta\omega(t_1)\rangle\propto\exp(-|t-t_1|/t_c),
\label{eq:noise}
\end{equation}
in which $t_c$ denotes the autocorrelation time.
To keep the models simple, we are choosing to add the noise to the core-oscillation frequency; however, we have also confirmed that adding colored noise to the external focusing frequency does not significantly change the results.

\begin{figure}
\includegraphics[width=15cm]{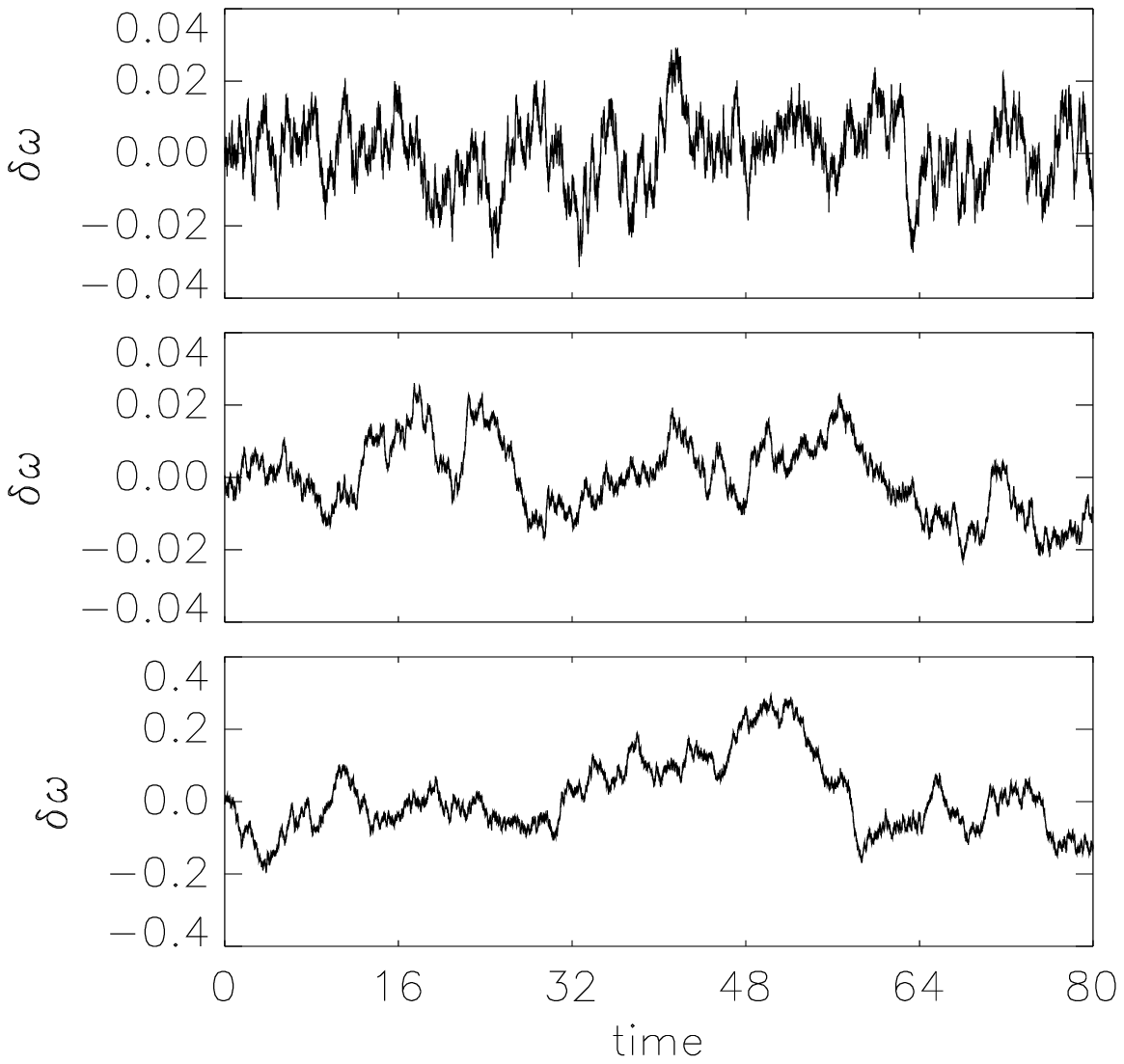}
\caption{\label{fig:noise}Example manifestations of colored noise along an orbit for $\langle|\delta\omega|\rangle=0.01$ and for which $t_c=1.5\tau$ (top) and $12\tau$ (center), and for $\langle|\delta\omega|\rangle=0.1$ with $t_c=12\tau$ (bottom).} 
\end{figure}

After generating a colored-noise signal using an algorithm first presented in Ref.~\cite{pogorelov99}, we calculate $\langle|\delta\omega|\rangle$ which becomes a measure of the noise strength.
The influence of noise on halo formation should in principle depend on its strength and its autocorrelation time.
For two choices of autocorrelation time, $t_c=1.5\tau$ and $12\tau$, $\tau$ denoting the orbital period of a typical halo particle, we investigated a broad range of strengths, specifically $10^{-5}\leq\langle|\delta\omega|\rangle\leq 1$, with the goal of ascertaining to what extent the results may be regarded as generic.
Manifestations of colored noise that a particle might see are illustrated in Fig.~\ref{fig:noise}, which is provided as an aid toward conceptualizing the physical meaning of the noise parameters.
Shown there are manifestations of noise for a fixed strength $\langle|\delta\omega|\rangle=0.01$ with $t_c=1.5\tau$ and $12\tau$, and for $\langle|\delta\omega|\rangle=0.1$ with $t_c=12\tau$.

In a real beam each individual particle will have its own distinct initial conditions and thus experience a manifestation of the noise that differs from that seen by each of the other particles.
For example, in the axisymmetric Model I, each particle initially occupying a thin annulus centered at radius $x(0)$ will experience noise differing from that seen by each of the other particles initially in that annulus because the particles start at different angular coordinates.
The same is true for particles initially occupying a spherical shell centered on radius $r(0)$ in Model II.
Accordingly, we adopted a `survey strategy'.
Upon choosing initial conditions $x(0)$ and $r(0)$ for Eqs.~(\ref{eq:wangler}) and (\ref{eq:TE}), respectively, and for a specific choice of noise parameters, we sequentially computed 10,000 orbits, each experiencing its own random manifestation of the colored noise, and we catalogued the maximum amplitudes of these orbits.
We set the initial conditions of the orbit in Model I at $x(0)=1.20$, $\dot{x}(0)=0$, and in Model II at $r(0)=1.23$, $\dot{r}(0)=0$.
In the unperturbed TE sphere of Model II, and for realistic proton beam parameters, there are $\sim 4\times 10^9$ particles per bunch, i.e., $\sim 0.6$ nC~\cite{kandrup03}.
There are $\sim 3\times 10^4$ particles in the range $r=1.23\pm (0.5\times 10^{-4})$, a thin spherical shell centered on $r(0)$ and located well into the Debye tail of the bunch.
Accordingly, the chosen sample size is realistic.

\begin{figure}
\includegraphics[width=15cm]{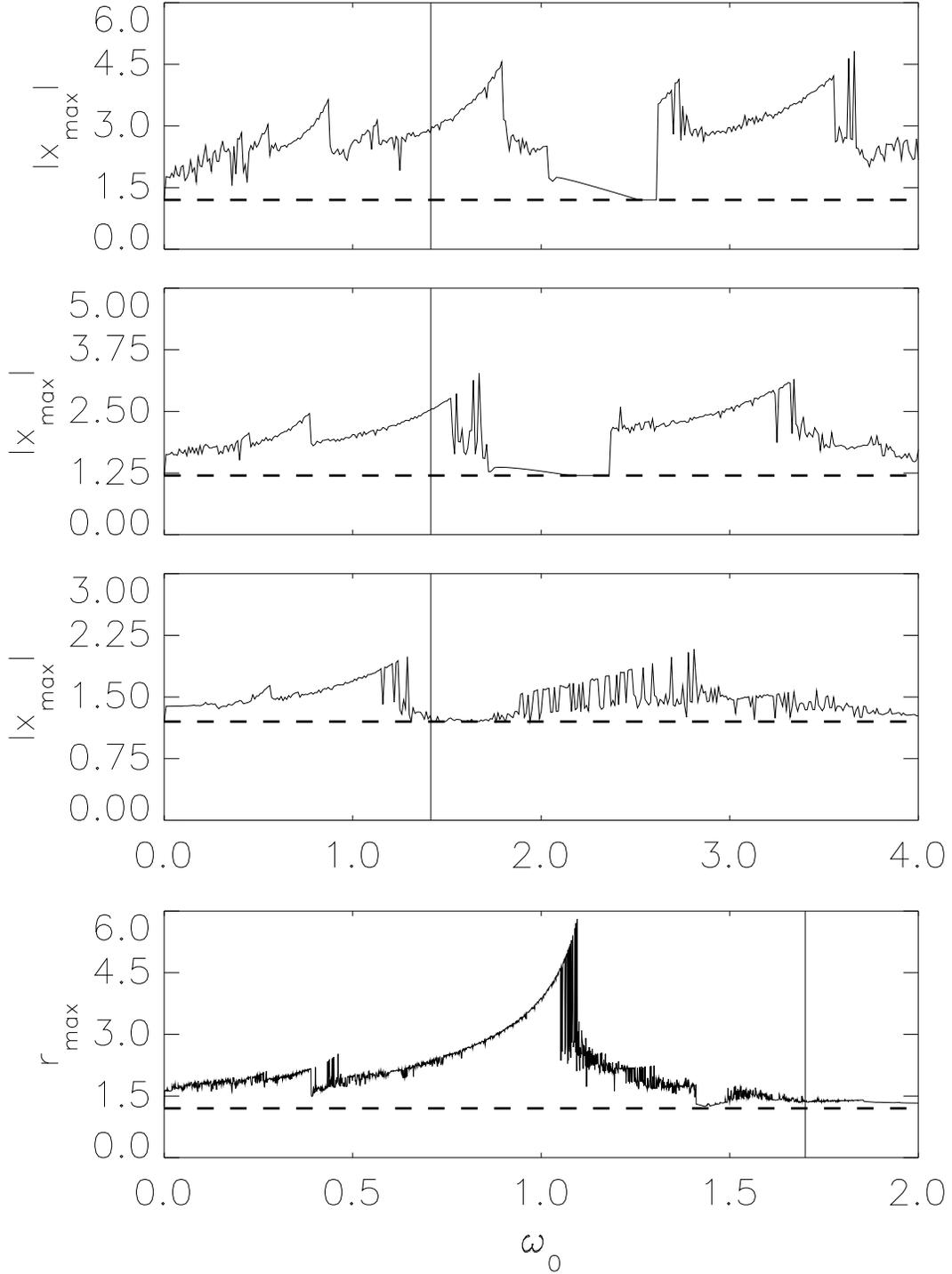}
\caption{\label{fig:nonoise}(top 3 panels) Maximum orbital amplitude vs. core-oscillation frequency $\omega_0$ with zero noise in Model I for mismatch parameters $M=1.5,1.3,1.1$; the horizontal line denotes the initial condition $x(0)=1.20$ and the vertical line denotes the frequency choice $\omega_0=\sqrt{2}$.
(bottom panel) Same for Model II for perturbation parameter $\mu=0.5$; the horizontal line denotes the initial condition $r(0)=1.23$ and the vertical line denotes the frequency choice $\omega_0=1.7$.}
\end{figure}

For Model I, we examine three values of the mismatch parameter: $M=1.5,\;1.3,\;1.1$.
Orbits are computed from Eq.~(\ref{eq:wangler}) first without, then with, the noise of Eq.~(\ref{eq:noise}) using a variable-time-step integrator.
For zero noise, the maximum orbital amplitude $|x_{max}|$ does have a hard upper bound in keeping with parametric-resonance arguments, and the upper bound depends on the core-oscillation frequency $\omega_0$.
As Fig.~\ref{fig:nonoise} shows, the particle can reach relatively large amplitudes for a wide range of frequencies $\omega_0$, a consideration that can be important in the context of higher-order space-charge modes and harmonics.
For nonzero colored noise, we present results for which the core-oscillation frequency is fixed at $\omega_0=\sqrt{2}$, the value obtained from the linearized equation for the motion of the beam envelope.
By design, then, Model I is a direct generalization of the particle-core model introduced in Ref.~\cite{oconnell93}; we found that different choices of $\omega_0$ do not change the essential findings.

\begin{figure}
\includegraphics[width=15cm]{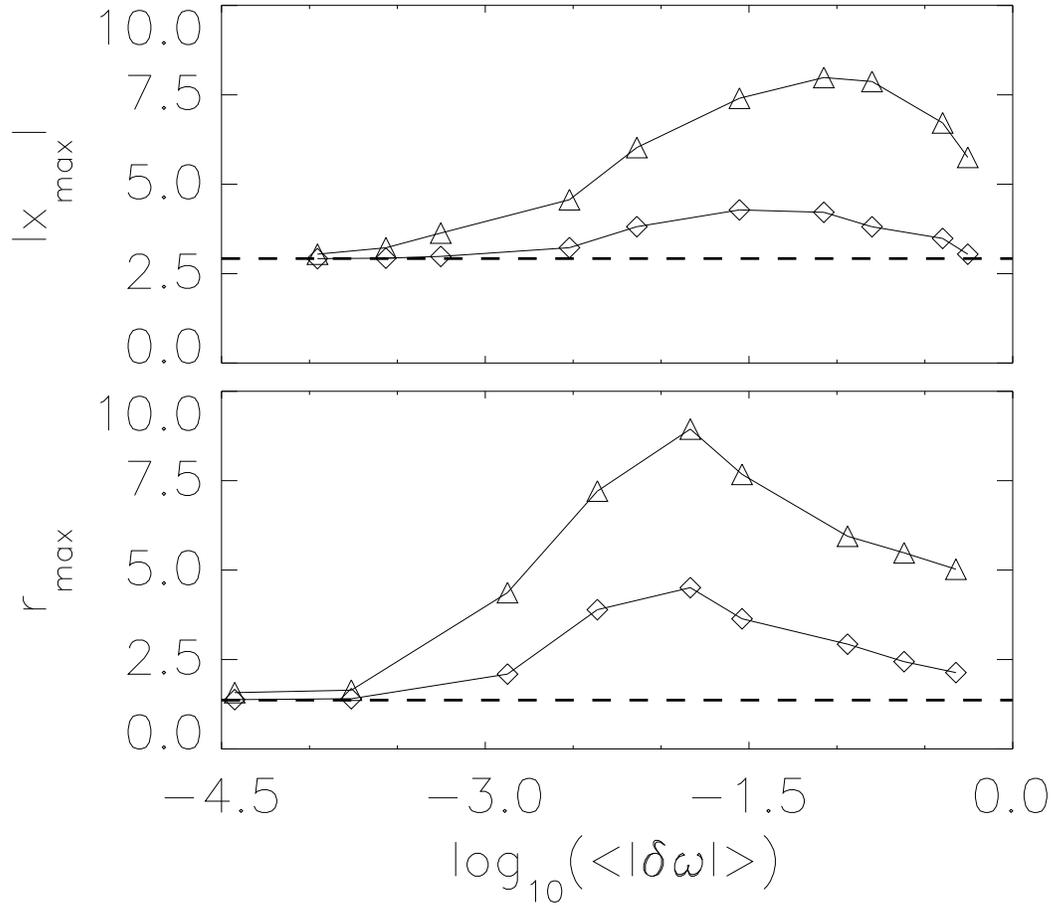}
\caption{\label{fig:statistics}(top) For the 10,000-particle sample in the potential of Model I with mismatch parameter $M=1.5$ and $t_c=12\tau$, largest amplitude reached by any particle (triangles) and mean maximum amplitude of all particles (diamonds) vs. $\langle|\delta\omega|\rangle$; the dashed line denotes $|x_{max}|$ for zero noise.
(bottom) Same for Model II with perturbation parameter $\mu=0.5$.}
\end{figure}

For specified noise parameters, we consider the one particle out of the sample of 10,000 that reaches the largest amplitude during the integration time of 80$\tau$, a time that is representative of the transit time through a 1 GeV proton linac.
Results for $M=1.5$ and fixed $t_c=12\tau$ are provided in the top panel of Fig.~\ref{fig:statistics}, in which $|x_{max}|$ versus $\langle|\delta\omega|\rangle$ is plotted; results for $M=1.3,\;1.1$ are qualitatively similar.
The figure also shows the average $|x_{max}|$ reached by particles in the sample.
One sees that over a broad range of noise strengths, rare particles are ejected to larger amplitudes relative to the parametric resonance alone.
For example, a mere 1$\%$ fluctuation in the core-oscillation frequency more than doubles the maximum amplitude reached compared to the case of zero noise.
Interestingly, we found for $t_c=1.5\tau$ that the results are very similar.

For the analysis of Model II, we fix the perturbation parameter at $\mu=0.5$.
Just as in Model I, with zero noise the particle can reach relatively large amplitudes for a wide range of frequencies $\omega_0$, as Fig.~\ref{fig:nonoise} shows.
For specified noise parameters, we present in the bottom panel of Fig.~\ref{fig:statistics} results for which $\omega_0=1.7$, a completely arbitrary choice of driving frequency; different choices of $\omega_0$ do not change the essential findings.
Models I and II are distinctly much different, yet the results make clear that the influence of the noise on the maximum orbital amplitudes in these models is nearly identical.
This is a remarkable finding in that we constructed Model II {\it ad hoc}, with no predisposition toward matching the results of Model I.
Accordingly, the influence of colored noise on particle orbits, and in particular its role in generating large distended halos in time-dependent potentials, appears to be generic.
Moreover, the collection of findings suggests that the formation of these halos is not particularly sensitive to details in either the governing potential or the noise.

\begin{figure}
\includegraphics[width=15cm]{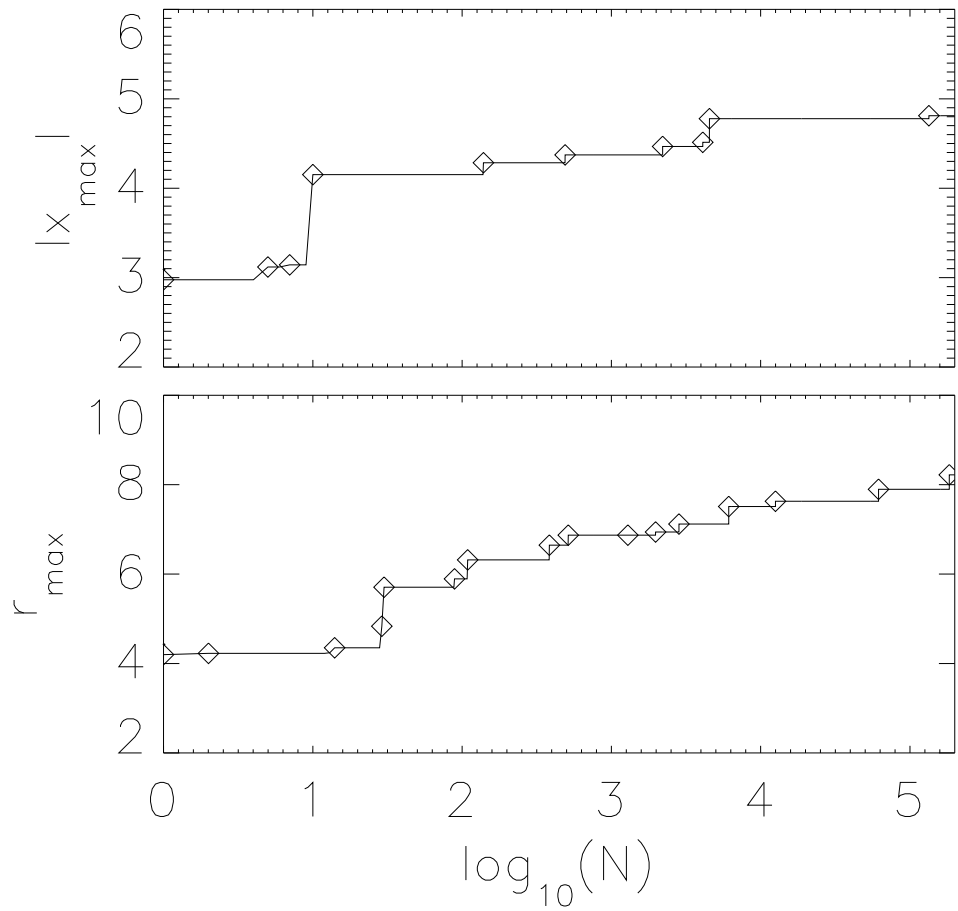}
\caption{\label{fig:samplesize}Maximum halo dimension vs. sample size $N$ for Model I (top) and Model II (bottom).}
\end{figure}

If the number of particles in the sample is increased with all else being the same, then the largest amplitude reached by the single special particle increases.
As Fig.~\ref{fig:samplesize} indicates, once the sample size is sufficiently large, the maximum amplitude grows quasi-logarithmically with increasing sample size.
Only in this restricted sense, and for fixed noise parameters and fixed integration time, may it safely be said that there is an upper bound to the halo dimension.
This particular point has actually been observed, but heretofore unexplained, in massive parallelized beam-dynamics simulations of an earlier design of the SNS linear accelerator that included a number of machine imperfections~\cite{qiang01}.
In runs involving $10^4$, then $10^5$, then $10^6$, then $10^7$ simulation particles, the maximum extent of the halo increased, but it seemed to approach a limiting value with runs above $10^8$ particles, a large number whose value generally depends on the details of the potential.
Inasmuch as these runs were self-consistent, the phenomenology they reflect is suggestive of the influence of increasingly fine resolution of details in the potential that are beyond the scope of a simple particle-core model.
They also exemplify that a large number of particles is needed to discern the impact of these details on halo formation and structure.

\begin{figure}
\includegraphics[width=15cm]{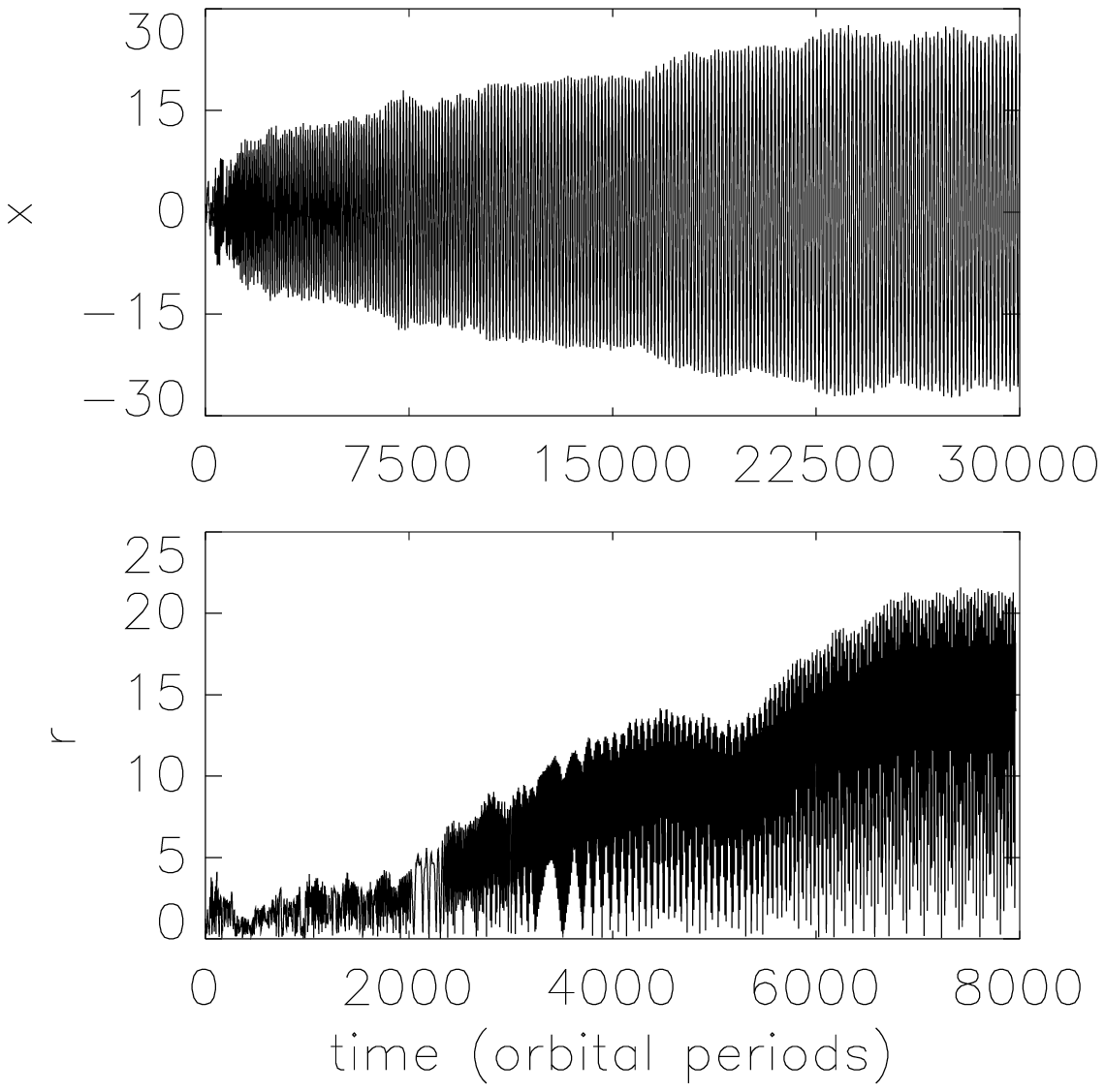}
\caption{\label{fig:longtime}Long-time evolution of a large-amplitude orbit given noise with $\langle|\delta\omega|\rangle=0.01$, $t_c=12\tau$ for Model I (top) and Model II (bottom).}
\end{figure}

If the integration time is extended indefinitely, as might be physically representative of a storage ring, for example, then there are statistically rare orbits that continue to grow to seemingly unlimited amplitudes.
Examples of such orbits in Models I and II appear in Fig.~\ref{fig:longtime}.
These long-time orbits exemplify that {\it there is in principle no upper bound to the halo amplitude in the presence of colored noise}.

It remains, of course, to explore further the extent to which this phenomenology applies in real machines.
Doing so will involve further simulations of beams in real beamlines; as we have seen, {\it machine imperfections will matter}.
One possibly fruitful approach is to extract the coarse-grained, time-dependent potential from the simulations and then add noise and pursue a statistical analysis of test particles in parallel to what we have done here.
Alternatively, the colored noise may be built directly into the simulation itself, although the simulation will then need to incorporate a sufficiently large number of particles to garnish enough statistics on the halo population.
A realistic manifestation of the colored noise would need to reflect the machine design, i.e., by properly including imperfections in the fields and hardware alignment, and details of the evolving space-charge potential such as a sufficiently detailed mode spectrum.
Of course, as the beam is accelerated and becomes relativistic, space charge and its attendant parametric resonance will become decreasingly important, and growth of the halo will thereby be curtailed.

As a relevant aside, we also analyzed this mechanism in the context of a self-gravitating stellar system for which environmental noise from surrounding galaxies will self-consistently influence the dynamics.
Specifically, we considered a perturbed Plummer model, a configuration for which the unperturbed collective potential scales as $(1+r^2/3)^{-1/2}$~\cite{binney}, and we applied the same procedure described herein for Model II.
Though it is a restoring force, gravity is so weak that, combined with the noise, only a relatively tiny oscillatory perturbation suffices to pump stars to very large amplitudes.
The main point, the generality of which is highlighted by the addition of this `gravitational' example, is that {\it colored noise combined with parametric resonance will drive a statistically small number of particles to much larger amplitudes than parametric resonance can do on its own}.
The formation of distended halos is thus a general byproduct of collective relaxation of nonequilibrium Coulomb systems.

This work was supported by the Department of Education under Grant G1A62056.

\newpage

\newpage

\newpage

\newpage

\newpage


\begin{thebibliography}{99}
\bibitem{jameson96}R. Jameson, Fus. Eng. Design {\bf 32-33}, 149 (1996).
\bibitem{sns03}Spallation Neutron Source Report No. 100000000-PL0001-R09, May 2003 (unpublished).
\bibitem{yjchen}Y.-J. Chen, {\it et al.}, in {\it Proceedings of the 1991 Particle Accelerator Conference}, edited by L. Lazema and J. Chew (IEEE, Piscataway, NJ, 1991), p. 3100.
\bibitem{oconnell93} J.S. O'Connell, T.P. Wangler, R.S. Mills, and K.R. Crandell, in {\it Proceedings of the 1993 Particle Accelerator Conference}, edited by S.T. Corneliussen (IEEE, Piscataway, NJ, 1993), p. 3657.
\bibitem{gluckstern94} R.L. Gluckstern, Phys. Rev. Lett. {\bf 73}, 1247 (1994).
\bibitem{okamoto97}H. Okamoto and M. Ikegami, Phys. Rev. E {\bf 55}, 4694 (1997).
\bibitem{gluckstern98}R. L. Gluckstern, A. V. Fedotov, S. Kurennoy, and R. Ryne, Phys. Rev. E {\bf 58}, 4977 (1998).
\bibitem{ikegami99}M. Ikegami, S. Machida, and T. Uesugi, Phys. Rev. ST Accel. Beams {\bf 2}, 124201 (1999).
\bibitem{ikegami299}M. Ikegami, Phys. Rev. E {\bf 59}, 2330 (1999).
\bibitem{qiang00}J. Qiang and R. Ryne, Phys. Rev. ST Accel. Beams {\bf 3}, 064201 (2000).
\bibitem{wang00}T.-S. F. Wang, Phys. Rev. E {\bf 61}, 855 (2000).
\bibitem{jeon02}D. Jeon, {\it et al.}, Phys. Rev. ST Accel. Beams {\bf 5}, 094201 (2002).
\bibitem{allen02}C.K. Allen, {\it et al.}, Phys. Rev. Lett. {\bf 89}, 214802 (2002).
\bibitem{bohn03}C.L. Bohn and I.V. Sideris, Phys. Rev. ST Accel. Beams {\bf 6}, 034203(2003); the unperturbed potential of Model II is the spherically symmetric `case 5' potential of this reference.
In the dimensionless units of Eq.~(\ref{eq:TE}), a representative value for the size of the matched beam is $R=10$; thus we divide all values of $r(t)$ for Model II by 10 to facilitate comparing results with those of Model I.
\bibitem{pogorelov99}I.V. Pogorelov and H.E. Kandrup, Phys. Rev. E {\bf 60}, 1567 (1999).
\bibitem{kandrup03}H.E. Kandrup, I.V. Sideris, and C.L. Bohn, Phys. Rev. ST Accel. Beams (submitted).
\bibitem{qiang01}J. Qiang, {\it et al.}, Nucl. Instrum. Methods {\bf A457}, 1 (2001).
\bibitem{binney}J. Binney and S. Tremaine, {\it Galactic Dynamics} (Princeton Univ. Press, Princeton, 1987), pp. 223-225.
\end{thebibliography}
\end{document}